\documentclass[aps,pre,epsfig,floats,twocolumn,superscriptaddress,amssymb,amsmath,floatfix,showpacs]{revtex4}
\usepackage{graphicx}
\usepackage{bm}  
\usepackage{amssymb}
\usepackage{color}
\usepackage{amscd}

\begin{document}
\title{Non-Markovianity by Quantum Loss}

\author{S. Haseli}

\author{S. Salimi}
\email{shsalimi@uok.ac.ir}
\affiliation{Department of Physics, University of Kurdistan, P.O.Box 66177-15175 , Sanandaj, Iran}

\date{\today}

\begin{abstract}
In the study of open quantum systems, information exchange between system and its surrounding environment plays an eminent and important role in analysing the dynamics of open quantum system. In this work, by making use of the quantum information theory and intrinsic properties such as \emph{entropy exchange}, \emph{coherent information}  and using the notion of \emph{quantum loss} as a criterion of the amount of lost information, we will propose a new witness, based on information exchange, to detect non-Markovianity. Also a measure for determining the degree of non-Markovianity, will be introduced by using our witness. The characteristic of non-Markovianity is clarified by means of our witness, and we emphasize that this measure is constructed based on the loss of information or in other word the rate of \emph{quantum loss} in the environment. It is defined in term of reducing correlation between system and ancillary. Actually, our focus is on the information which be existed in the environment and it has been entered to the environment due to its interaction with the system. Remarkably, due to choosing the situation which the "system +ancillary" in maximal entangled pure state, optimization procedure does not need in calculation of our measure, such that the degree of non-Markovianity is computed analytically by straightforward calculations.

\end{abstract}
\maketitle
\section{Introduction}
The study of open quantum systems plays an important and fundamental role in many applications of quantum information theory \cite{1}. Since isolation of quantum system from its surrounding environment is almost impossible, thus realistic quantum systems are open ones. We emphasize that, they exchange information with its surrounding environment. Surveying and focusing on this information exchange lead us to the concept of Markovian and non-Markovian dynamical processes. If the degree of freedom of environment is infinite or that, the coupling between system and its environment be weak, information flow from system to environment and the quantum dynamical process is Markovian. If it is comparable with the degree of freedom of the quantum system then the information flow back to system and dynamical process is non-Markovian and memory effects are revealed. If the rate of the flow of the information from system to environment can be calculated then we can define the type of the quantum dynamical process. Several witnesses have been provided to detect non-Markovianity of dynamical process and parallel to these witnesses various measures were specified in order to identify the degree of non-Markovianity. For example, Wolf et al. proposed their measure by using the semigroup property of the Markovian dynamical maps \cite{2}, Rivas et al. introduced their measure based on divisibility of Markovian dynamical maps \cite{3}, these two do not include the optimization procedure. However, several measures were presented from information exchange point of view. For instances, Breuer et al. provided a measure based on back flow of information by making use of the distinguishability notion \cite{4}, Luo et al. proposed a non-Markovianity measure via correlation based on mutual information \cite{5}. As can be seen in the above mention approach to determine the non-Markovian process we need to optimization procedure over all initial state. We have various and relevant , but conceptually different, description for non-Markovianity. Our motivation in writing this work has been to try to introduce a new measure in order to detect non-Markovianity and specify the degree of it which has a comprehensive physical interpretation and mathematically computable in a simple way. For this purpose, we use the notion of "\emph{quantum loss}" \cite{6} as the criteria for determining the value of lost information during the process. If we have the dynamics of lost information then we can specify the type of dynamical process. In this work, we  use our measure for popular examples and  will show that the obtained results are same results which obtain by other  measures based on information flow, but our measure have a advantage than before measures does not require optimization procedure due to choosing the situation which the "system +ancillary" in maximal entangled pure state.
\\ The work is organized as follows. In Sec.II we briefly review those appearances of quantum information theory, and introduce our model and the notions of "\emph{coherent information}", "\emph{quantum loss}" and "\emph{quantum noise}". In Sec.III we introduce our measure based on lost information by means of the quantum loss notion. In Sec.IV we use our measure for popular examples. Finally in Sec.V our results will be summarized in conclusions.
\section{MODEL}
Quantum information has special properties different from classical one. Quantum state of an open quantum system lost its coherence due to interacting with environment. This interaction induces an exchange of information between system and its surrounding environment, so we expect that, one can obtain much information about open quantum system by using the entropy of open quantum systems and its properties. Quantum entropy was introduced by von Neumann as an extension of Gibbs entropy in classical statistical mechanic \cite{7}. Here, we use von Neumann entropy to investigate the properties of open quantum system during its evolution. The basis of modern information theory was based on using the definition of entropy by Shannon \cite{8}. For a quantum state $\rho^{AB}$, we can write the relation between von Neumann entropies as
\begin{equation}\label{a}
  S(\rho^{A|B})=S(\rho^{AB})-S(\rho^{B}).
\end{equation}
Similarly, mutual information is given by
\begin{equation}\label{b}
  I(\rho^{AB})=S(\rho^{A})+S(\rho^{B})-S(\rho^{AB}),
\end{equation}
mutual information is bounded
\begin{equation}\label{c}
  0\leqslant I(\rho^{AB})\leqslant 2 min \{S(\rho^{A}),S(\rho^{B})\}.
\end{equation}
For pure bipartite state $\rho^{AB}$ from Eq.(\ref{c}), and making use of this fact that, we can use the von Neumann entropy of reduce density matrix to measure the entanglement for pure state\cite{9,10}, mutual information is rewritten as
\begin{equation}\label{d}
  I(\rho^{AB})=I_{Q}(\rho^{AB})=2S(\rho^{A})=2S(\rho^{B}),
\end{equation}
where $I_{Q}$  was defined as mutual entanglement \cite{11}. We consider an open quantum system $\mathcal{S}$ with Hilbert
space $\mathcal{H}_{\mathcal{S}}$, and arbitrary density matrix $\rho^{\mathcal{S}}$ belong to, all
bounded linear operators acting on Hilbert space, $B(\mathcal{H}_{\mathcal{S}})$. Initial mixed state $\rho^{\mathcal{S}}$ is purified by using entanglement with an ancillary system $\mathcal{A}$ with Hilbert
space $\mathcal{H}_{\mathcal{A}}$, and  density matrix $\rho^{\mathcal{A}}$ belong to, all
bounded linear operators acting on Hilbert space, $B(\mathcal{H}_{A})$
\begin{equation}\label{e}
  |\Psi^{\mathcal{SA}}\rangle = \sum_{i}\sqrt{\lambda_{i}}|i,a_{i}\rangle,
\end{equation}
where $|a_{i}\rangle$'s are eigenstates of $\mathcal{A}$,  $\lambda_{i}$ and $|a_{i}\rangle$ are eigenvalues and eigenstates of $\mathcal{S}$ respectively, it can be achived via a Schmidt decomposition. Open quantum system interact with its environment with Hilbert space$\mathcal{H}_{\mathcal{E}}$ and density matrix $\rho^{\mathcal{E}}$. We supposed that the environment initially is in pure state. Ancillary system does not evolve during the evolution and $\rho^{\mathcal{SE}}$ evolves unitarily in time, so due to these facts the total state $\rho^{\mathcal{SAE}}$ is pure and remain pure in time. Total state evolve under unitary operation $U^{\mathcal{SE}}\otimes I^{\mathcal{A}}$,
\begin{equation}\label{f}
  \rho^{\mathcal{\acute{S}\acute{E}A}}=(U^{\mathcal{SE}}\otimes I^{\mathcal{A}})\rho^{\mathcal{SAE}}(U^{\mathcal{SE}}\otimes I^{\mathcal{A}})^{\dag}.
\end{equation}
This construction is summarized in Fig. (1).
\begin{figure}[h]
  \centering
  \includegraphics[width=9cm]{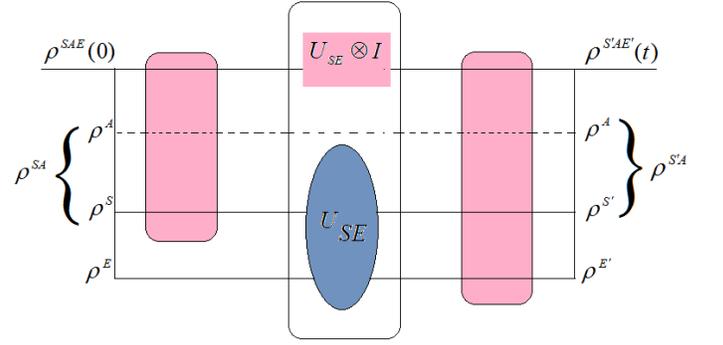}\\
  \caption{(Color online) The initial state of total quantum state $\rho^{\mathcal{SAE}}$ is generally in a pure state. The state of the ancilla $\mathcal{A}$ is chosen such that it purify $\rho^{\mathcal{S}}$. The state of combine state $\rho^{\mathcal{SAE}}$ remain pure in time due to unitary evolution of total system with unitary operation $U^{\mathcal{SE}}\otimes I^{\mathcal{A}}$.}\label{2}
\end{figure}
Befor the interaction with environment, due to the fact that initial state of composite "\emph{system + ancillary}" is in pure state, the entropy of ancillary state is the same with entropy of system state and is denoted by
\begin{equation}\label{g}
  S(\rho^{\mathcal{A}})=S(\rho^{\mathcal{S}})=-\sum_{i}\lambda_{i}\log_{2}\lambda_{i}.
\end{equation}
After the evolution the entropy of environment known as "\emph{ entropy exchange}" which is introduced by Schumacher \cite{5}
\begin{equation}\label{h}
  S_{e}=S(\rho^{\mathcal{\acute{E}}})=S(\rho^{\acute{\mathcal{S}}\mathcal{A}})=-Tr[\rho^{\acute{\mathcal{S}}\mathcal{A}}\log_{2} \rho^{\acute{\mathcal{S}}\mathcal{A}}].
\end{equation}
Entropy exchange determines the measure of information exchanged between system $\mathcal{S}$ and environment $\mathcal{E}$ during the quantum evolution, which is the von Neumann entropy of final state $\rho^{\acute{\mathcal{S}}\mathcal{A}}$. Entropy exchange is a natural properties of $\mathcal{S}$, it is only depend on $\rho^{\mathcal{S}}$  and dynamical map $\Lambda^{\mathcal{S}}$
\begin{equation}\label{i}
  \rho^{\mathcal{A\acute{S}}}=[\mathcal{I^{\mathcal{A}}}\otimes \Lambda^{\mathcal{S}}] \rho^{\mathcal{AS}}.
\end{equation}
If interaction between system and environment does not exist then system is close and its evolution is unitary, so the joint state $\rho^{\mathcal{SA}}$ remain pure in time. In this case entropy exchange is equal to zero.
Coherent information $I_{c}$ is an another natural quantity, which is given by \cite{13}
\begin{equation}\label{j}
  I_{c}=S(\rho^{\mathcal{\acute{S}}})-S(\rho^{\mathcal{A\acute{S}}})=S(\rho^{\mathcal{\acute{S}}})-S_{e}.
\end{equation}
By using coherent information the "quantum loss" is obtained as
\begin{equation}\label{k}
  L_{Q}=S(\rho^{\mathcal{A}}:\rho^{\mathcal{\acute{E}}}|\rho^{\mathcal{S}})=S(\rho^{\mathcal{S}})-I_{c},
\end{equation}
quantum loss play an important role in quantum error correction \cite{5}. Following triangle inequality holds for quantum loss \cite{11,14}
\begin{equation}\label{l}
  0\leq L_{Q}\leq 2min\{S(\rho^{\mathcal{S}}),S_{e}\}.
\end{equation}
\begin{figure}[hb]
  \centering
  \includegraphics[width=9cm]{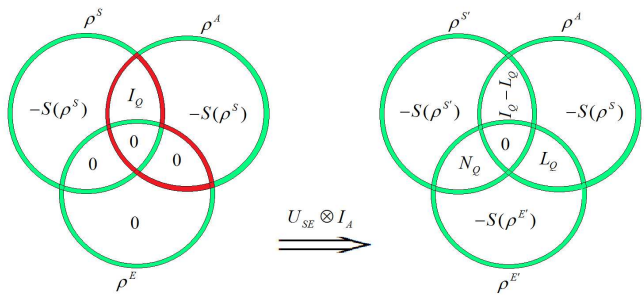}\\
  \caption{(Color online)Unitary evolution of total pure system $|\Psi^{\mathcal{SAE}}\rangle$. Ancillary system does not evolve during the unitary transformation $U^{\mathcal{SE}}\otimes I^{\mathcal{A}}$ and so there is no exchange across the red solid line on the left entropy venn diagram. On the right venn diagram $N_{Q}=2S(\rho^{\mathcal{E}})-L_{Q}$.}\label{2}
\end{figure}
From Fig.(2) we find that  "\emph{quantum loss}" $L_{Q}$ indicates the loss of information from system to environment or back flow of information from environment to system  in the quantum evolution, we will show these notions in the following section analytically. Initial mutual entanglement for pure state $I_{Q}=2S(\rho^{\mathcal{S}})$ has the information dimension. By interaction of the system with environment, it is converted to
\begin{equation}\label{m}
  \acute{I}_{Q}=I_{Q}-L_{Q}.
\end{equation}
Note that ternary mutual information ,i.e. the center of venn diagram in Fig.2, due to monogamy of mutual information must satisfy inequality which is given by
\begin{equation}\label{h1}
  I(\rho^{\mathcal{S}}:\rho^{\mathcal{A}}:\rho^{\mathcal{E}})\leq 0, \quad I(\rho^{\mathcal{\acute{S}}}:\rho^{\mathcal{A}}:\rho^{\mathcal{\acute{E}}})\leq 0.
\end{equation}
Here we assume that the total state $\rho^{SAE}$ in pure state, so the ternary mutual information(the center of the venn diagram in Fig.2) is equal to zero \cite{15} .
\section{NON-MARKOVIANITY VIA QUANTUM LOSS}
Supposed that $\Lambda^{\mathcal{S}}=\{\Lambda_{t}\}$ be a quantum dynamical map, which is defined by a set of linear and completely positive trace preserving operations$\Lambda_{t}$.  $\Lambda_{t}$ act on state space $B(\mathcal{H_{\mathcal{S}}})$ of the system. "Ancillary + system" Hilbert space is $H^{\mathcal{S}}\otimes H^{\mathcal{A}}$.
If $\{\Lambda_{t}\}$ be a Markovian dynamical map i.e. for $r\leqslant t$
\begin{equation}\label{n}
 \Lambda_{t}=\Lambda_{t,r}\Lambda_{r}.
\end{equation}
For the state  of the combined "ancillary + system"  Put $\rho^{\mathcal{SA}}(t)=(\Lambda_{t}\otimes I)\rho^{\mathcal{SA}}(0)$, note that ancillary system does not evolve. From the visual point of view in Fig.2 and Eq.(\ref{m}), we observe that, if the "\emph{quantum loss}" $L_{Q}$ increase continuously, i.e. $\frac{d}{dt}L_{Q}\geq 0$, the value of mutual entanglement, i.e. correlation between system and ancillary, decrease monotonically, and information flow from system to environment, so the dynamical process is Markovian. On the other hand if during the process the value of quantum loss decrease in some time intervals, i.e. $\frac{d}{dt}L_{Q}< 0$, then the value of mutual entanglement increase, and dynamical process is non-Markovian. Now we are in a position to define our witness for detecting non-Markovianity.

\subsection{NON-MARKOVIANITY WITNESS}
 The quantum dynamical map  $\Lambda_{t}$ is non-Markovian iff $\frac{d}{dt}L_{Q}(\rho^{\mathcal{SA}}(t))< 0$. In other words  violation of monotonically increasing  property of "quantum loss" under quantum dynamical process, i.e. $\frac{d}{dt}L_{Q}(\rho^{\mathcal{SA}}(t))\geq 0$, is necessary and sufficient condition for  non-Markovianity of dynamical maps. According to our witness, we find that in non-Markovian process the rate of the Von Neumann entropy of environment is less than the rate of the system's von Neumann entropy
\begin{equation}\label{eqe}
  \frac{d}{dt}S(\rho^{\mathcal{E}}(t))<\frac{d}{dt}S(\rho^{\mathcal{S}}(t)).
\end{equation}
Similar to pervious work on this fields by other authors, we can introduce a measure for determining the degree of non-Markovianity using this witness.

\subsection{NON-MARKOVIANITY MEASURE}
The above constructions
help us to make a natural measure for indicating the degree of non-Markovianity of quantum dynamical map $\Lambda_{t}$ from quantum loss dynamics
\begin{equation}\label{s}
  \mathcal{N}_{L_{Q}}(\Lambda_{t})=\int_{\frac{d}{dt}L_{Q}(\rho^{\mathcal{SA}}(t))< 0}\frac{d}{dt}L_{Q}(\rho^{\mathcal{SA}}(t)) dt.
\end{equation}
Integral is over all time intervals $t\in(a_{i},b_{i})$ in which $\frac{d}{dt}L_{Q}(\rho^{\mathcal{SA}}(t))< 0$ and $\rho^{\mathcal{SA}}(0)$ is maximally entangled pure state.

Mathematically, by taking the time derivative from both sides of the Eq.(\ref{m}) we have
\begin{equation}\label{r}
  \frac{d}{dt}I_{Q}(\rho^{\mathcal{SA}}(t))=-\frac{d}{dt}L_{Q}(\rho^{\mathcal{SA}}(t)),
\end{equation}
which means that, increasing the quantum loss is equivalent to a reduction of mutual entanglement and vice versa. In other meaning, we can say that the dynamical map is Markovian if $\frac{d}{dt}I_{Q}(\rho^{\mathcal{SA}}(t))\leq0$, and is non-Markovian iff $\frac{d}{dt}I_{Q}(\rho^{\mathcal{SA}}(t))>0$. From Eq.(\ref{d}) we observe that the mutual entanglement $I_{Q}((\rho^{\mathcal{SA}}(t))$ has the dimension of mutual information $I((\rho^{\mathcal{SA}}(t))$ for bipartite pure states. Thus, with regard to above considerations, we can reproduce the witness which were introduced by Luo et al. in Ref. \cite{5}. From their work, we have following relation for quantum mutual information of combined system "system + ancillary" during the quantum evolution
\begin{equation}\label{o}
 I(\rho^{\mathcal{SA}}(t))\leq I(\rho^{\mathcal{SA}}(r)),
\end{equation}
i.e. if $\{\Lambda_{t}\}$ be a Markovian quantum dynamical map then the quantum mutual information is a monotonically decreasing function of $t$, or in other word for any Markovian dynamical maps we have
\begin{equation}\label{p}
  \frac{d}{dt}I(\rho^{\mathcal{SA}}(t))\leq 0.
\end{equation}
Any violation of this monotonicity is a symptom of non-Markovianity.
\section{EXAMPLES}

{\bf \textit{Example 1:}}Consider pure dephasing of a single-qubit system
described by the following local generator
\begin{equation}\label{t}
 \dot{\rho}^{S}(t)= \mathcal{L}_{t}(\rho^{\mathcal{S}}(t))=\frac{\gamma(t)}{2}(\sigma_{z}\rho^{\mathcal{S}}(t)\sigma_{z}-\rho^{\mathcal{S}}(t)).
\end{equation}
 By solving this master equation we have
\begin{equation}\label{u}
\Lambda_{t}(\rho^{\mathcal{S}}(0))=\rho^{\mathcal{S}}(t)=\begin{pmatrix}
\rho_{11}(0) & \rho_{12}(0)e^{-\Gamma(t)} \\
\rho_{21}(0)e^{-\Gamma(t)} &\rho_{22}(0)
\end{pmatrix},
\end{equation}
where $\Gamma(t)=\int_{0}^{t}\gamma(\tau)d\tau$. Now we consider the ancillary system such that $\rho^{\mathcal{SA}}(0)=|\Psi^{\mathcal{SA}}(0)\rangle\langle \Psi^{\mathcal{SA}}(0)|$, where $
  |\Psi^{\mathcal{SA}}(0)\rangle = \frac{1}{\sqrt{2}}(|0^{\mathcal{S}}0^{\mathcal{A}}\rangle + |1^{\mathcal{S}}1^{\mathcal{A}}\rangle)$. Straightforward calculation lead to density matrix $\rho^{\mathcal{SA}}(t)$. From Eq.(\ref{k}), quantum loss for evolved "ancillary + system" composition can be obtained as
\begin{equation}\label{v}
 \begin{split}
L_{Q}(\rho^{\mathcal{SA}}(t))= & -\frac{1+\exp(-\Gamma(t))}{2}\log_{2}\frac{1+\exp(-\Gamma(t))}{2}- \\
     & -\frac{1-\exp(-\Gamma(t))}{2}\log_{2}\frac{1-\exp(-\Gamma(t))}{2}.
 \end{split}
 \end{equation}

By considering the Markovianity condition $\frac{d}{dt}L_{Q}(\rho^{\mathcal{SA}}(t))\geq 0$ after straightforward calculations we find that quantum dynamical map is Markovian if $\gamma(t)\geq 0$. Remarkably, from $\frac{d}{dt}L_{Q}(\rho^{\mathcal{SA}}(t))< 0$ quantum dynamical map is non-Markovian iff $\gamma(t)< 0$.
\vskip 1cm
 {\bf \textit{Example 2:}}As an another example we consider the decay of a two-level system into a bosonic environment. The total Hamiltonian is given by
 \begin{equation}\label{w}
   H_{t}=\omega_{0}\sigma_{+}\sigma_{-}\otimes I + I \otimes \sum_{k}\omega_{k}b_{k}^{\dag}b_{k}+\sum_{k}(g_{k}\sigma_{+}\otimes b_{k}+g_{k}^{\ast}\sigma_{-}\otimes b_{k}^{\dag}).
 \end{equation}
 We consider the state of the environment in ground state i.e $\rho^{\mathcal{E}}=|0\rangle\langle0|$. The dynamics of single qubit is described by the following time-local master equation
 \begin{equation}\label{x}
 \begin{split}
 \dot{\rho}^{S}(t)=&\mathcal{L}_{t}\rho^{S}(t)=-\frac{i}{2}S(t)[\sigma_{+}\sigma_{-},\rho^{S}(t)]  \\
     & +\frac{\gamma(t)}{2}(\sigma_{-}\rho^{S}(t)\sigma_{+}-\{\sigma_{+}\sigma_{-},\rho^{S}(t)\}),
 \end{split}
 \end{equation}
 where $S(t)=-2\mathfrak{I} \mathfrak{m} (\frac{\dot{G}(t)}{G(t)})$ and $\gamma(t)=-2\mathfrak{R}\mathfrak{e}(\frac{\dot{G}(t)}{G(t)})$. If the spectral density has a Lorentzian form $J(\omega)=\frac{1}{2\pi}\frac{\gamma_{0}\lambda^{2}}{(\omega_{0}-\omega)^{2}+\lambda^{2}}$,  $\lambda$
defines the spectral width of the coupling, which is connected to the reservoir
correlation time of the environment by $\tau_{B} = \lambda^{-1}$ and $\gamma_{0}$ is coupling constant to environment, which is related to the time scale of the system $\tau_{R}$ by the relation $\tau_{R}=\gamma_{0}^{-1}$. In this case $S(t)=0$ and the function $G(t)$ is derived as
\begin{equation}\label{y}
 G(t)=e^{\frac{-\lambda t}{2}}[\cosh(\frac{d t }{2})+\frac{\lambda}{d}\sinh(\frac{d t }{2})],
\end{equation}
where $d=\sqrt{\lambda^{2}-2\gamma_{0}\lambda}$. By solving the relevant master equation in Eq.(\ref{x}), density matrix in time $ t\geq 0$ defined as
\begin{equation}\label{z}
\rho^{\mathcal{S}}(t)=\begin{pmatrix}
1-|G(t)|^{2}\rho_{22}(0) & \rho_{12}(0)G(t) \\
\rho_{12}^{\ast}(0)G^{\ast}(t) &|G(t)|^{2}\rho_{22}(0)
\end{pmatrix}.
\end{equation}
Let us consider the ancillary, which is entangled with single qubit system such that $\rho^{\mathcal{SA}}(0)=|\Psi^{\mathcal{SA}}(0)\rangle\langle \Psi^{\mathcal{SA}}(0)|$, where $
  |\Psi^{\mathcal{SA}}(0)\rangle = \frac{1}{\sqrt{2}}(|0^{\mathcal{S}}0^{\mathcal{A}}\rangle + |1^{\mathcal{S}}1^{\mathcal{A}}\rangle)$. By extending the action of dynamical map to two-qubit systems, in recognition of, and with knowledge about the fact that ancillary system does not evolve, transformed density matrix is derived as
   \begin{equation}\label{a1}
\rho^{\mathcal{SA}}(t)=\frac{1}{2}\begin{pmatrix}
 1&0&0& G^{\ast}(t) \\
0 &1-|G(t)|^{2}&0&0\\
0&0&0&0\\
G(t)&0&0&G(t)
\end{pmatrix}.
\end{equation}
According to definition, quantum loss can be obtained as
\begin{equation}\label{b1}
 \begin{split}
 L_{Q}(\rho^{\mathcal{SA}}(t))=&1+\frac{2-|G(t)|^{2}}{2}\log_{2}\frac{2-|G(t)|^{2}}{2}+  \\
     & +\frac{|G(t)|^{2}}{2}\log_{2}\frac{|G(t)|^{2}}{2}- \\
     &-\frac{1-|G(t)|^{2}}{2}\log_{2}\frac{1-|G(t)|^{2}}{2}-\\
     & -\frac{1+|G(t)|^{2}}{2}\log_{2}\frac{1+|G(t)|^{2}}{2}.
 \end{split}
\end{equation}
From $\frac{d}{dt}L_{Q}(\rho^{\mathcal{SA}}(t))\geq 0$ for Markovian dynamical maps, we find dynamical map is Markovian if $\frac{d|G(t)|}{dt}\leq 0$. In other word, the map is Markovian when $|G(t)|$ be a monotonically decreasing function of $t>0$. Any violation from the condition which is mentioned, i.e.  establishment of $\frac{d}{dt}L_{Q}(\rho^{\mathcal{SA}}(t))< 0$, leads to non-Markovian dynamical map, in other word the dynamical map is non-Markovian iff $\frac{d}{dt}|G(t)|>0$
\vskip 1cm
 {\bf \textit{Example 3:}} Let us consider the random unitary single qubit dynamics by following time-dependent generator
 \begin{equation}\label{c1}
    \mathcal{L}_{t}(\rho^{\mathcal{S}}(t))=\frac{1}{2}\sum_{k=1}^{3}\gamma_{k}(t)[\sigma_{k}\rho^{\mathcal{S}}(t)\sigma_{k}-\rho^{\mathcal{S}}(t)].
 \end{equation}
 By solving this master equation we find that
 \begin{equation}\label{d1}
 \begin{split}
   \rho^{\mathcal{S}}_{11}(t)= & \frac{1}{2}[(\rho_{11}(0)-\rho_{22}(0))\mathcal{F}(t)+1], \\
   \rho^{\mathcal{S}}_{22}(t)=  & \frac{1}{2}[-(\rho_{11}(0)-\rho_{22}(0))\mathcal{F}(t)+1],\\
    \rho^{\mathcal{S}}_{12}(t)=&\frac{1}{2}[(\rho_{12}(0)+\rho_{21}(0))\mathcal{G}(t)+(\rho_{12}(0)-\rho_{21}(0))\mathcal{H}(t)],\\
    \rho^{\mathcal{S}}_{21}(t)=& \frac{1}{2}[(\rho_{12}(0)+\rho_{21}(0))\mathcal{G}(t)-(\rho_{12}(0)-\rho_{21}(0))\mathcal{H}(t)],
 \end{split}
\end{equation}
where $\mathcal{F}(t)=e^{-(\Gamma_{1}(t)+\Gamma_{2}(t))}$, $\mathcal{G}(t)=e^{-(\Gamma_{2}(t)+\Gamma_{3}(t))}$, $\mathcal{H}(t)=e^{-(\Gamma_{1}(t)+\Gamma_{3}(t))}$, and $\Gamma_{k}(t)=\int_{0}^{t} \gamma_{k}(\tau)d\tau$. In a similar way to two pervious example and choosing the same, initial state "system + ancillary" composition, with them, we can extend the action of dynamical map to two qubit map as
\begin{equation}\label{e1}
\begin{array}{ccc}
\rho^{\mathcal{SA}}(t)=(\Lambda_{t}^S\otimes I^{\mathcal{A}})\rho^{\mathcal{SA}}(0)= &  &  \\
 &  &  \\
\frac{1}{4}\begin{pmatrix}
 1+\mathcal{F}(t)&0&0& \mathcal{G}(t)+\mathcal{H}(t) \\
0 &1- \mathcal{F}(t)&\mathcal{G}(t)-\mathcal{H}(t)&0\\
0&\mathcal{G}(t)-\mathcal{H}(t)&1- \mathcal{F}(t)&0\\
\mathcal{G}(t)+\mathcal{H}(t)&0&0&1+\mathcal{F}(t)
\end{pmatrix}. &  &
\end{array}
\end{equation}
From the definition, in this case, we observe that  "quantum loss" is equal to  von Neumann entropy of $\rho^{\mathcal{SA}}(t)$ and is obtained as
\begin{equation}\label{f1}
  L_{Q}(\rho^{\mathcal{SA}}(t))=-\sum_{i=1}^{4}\lambda_{i}(t)\log_{2}\lambda_{i}(t),
\end{equation}
where $\lambda_{i}(t)$'s are eigenvalues of  $\rho^{\mathcal{SA}}(t)$. Similar to the earlier examples, Straightforward calculations, lead to this fact that the dynamical map is Markovian if three following inequality be satisfied simultaneously
\begin{equation}\label{g1}
 \begin{split}
 &\\\gamma_{1}(t)+\gamma_{2}(t)\geq 0,&\\
 &\\
  \gamma_{1}(t)+\gamma_{3}(t)\geq 0,   & \\
                &\\
 \gamma_{2}(t)+\gamma_{3}(t)\geq 0,& \\
 \end{split}
\end{equation}

the quantum dynamical map is non-Markovian iff one of these three inequality does not  exist.
By considering the final result of Ex.(1), Ex.(2) and Ex.(3) we conclude that, these results are the same with others that are obtained by using different non-Markovianity measures which are based on information flow between system and environment, such as Breuer et al. \cite{4} and Luo et al. \cite{5} measures.
\section{CONCLUSIONS}
In this paper a new measure for non-Markvianity have been provided by bringing together the notions of "\emph{quantum loss}", "\emph{entropy exchange}" and "\emph{coherent information}". Our measure is constructed based on information exchange between system and environment. In order to construct our measure we use the physical interpretation of "\emph{quantum loss}", which it can be interpreted as the lost information of open quantum system during its evolution. If the rate of "\emph{quantum loss}" is positive during the process $\frac{d}{dt}L_{Q}(\rho^{\mathcal{SA}}(t))\geq 0$, i.e. $L_{Q}(\rho^{\mathcal{SA}}(t))$ is monotonically increasing function of $t\geq0$, then the information flow from system to environment continuously and quantum dynamical map is Markovian. Any violation from Markovian condition is a sign of non-Markovianity. In other word the dunamical map is non-Markovian iff $\frac{d}{dt}L_{Q}(\rho^{\mathcal{SA}}(t))< 0$. In this letter we examine three popular example(dephasing, amplitude damping and depolarizing dynamics) by our measure. The results show that, this measure is consistent with the other measures which is constructed based on information exchange. Advantage of this criterion in comparison with other criteria is the simplicity of mathematical calculations, remarkably, optimization procedure does not exist in our measure. From the perspective of physical perception, the eminent character of non-Markovianity, i.e. back flow of information, is described by our measure more clearly. This description has been achieved via investigation of the information which is exist in the environment, and it has been entered to environment due to its interaction with the system. Also, we conclude that the Luo measure in Ref.\cite{5} can be reproduced by our measure.

\end{document}